# A Method for Ontology-based Architecture Reconstruction of Computing Platforms

Seyyedeh Atefeh Musavi,and Mahmoud Reza Hashemi
*School of Electrical and Computer Engineering,*
*College of Engineering,*
*University of Tehran, Tehran, Iran.*
*amusavi,rhashemi@ut.ac.ir*

**Abstract**—Today's ubiquitous computing ecosystem involves various kinds of hardware and software technologies for different computing environments. As the result, computing systems can be seen as integrated system of hardware and software systems. Realizing such complex systems is crucial for providing safety, security, and maintenance. This is while the characterization of computing systems is not possible without a systematic procedure for enumerating different components and their structural/behavioral relationships. Architecture Reconstruction (AR) is a practice defined in the domain of software engineering for the realization of a specific software component. However, it is not applicable to a whole system (including HW/SW). Inspired by Symphony AR framework, we have proposed a generalized method to reconstruct the architecture of a computing platform at HW/SW boundary. In order to cover diverge set of existing HW/SW technologies, our method uses an ontology-based approach to handle these complexities. Due to the lack of a comprehensive accurate ontology in the literature, we have developed our own ontology -called PLATOnt- which is shown to be more effective by ONTOQA evaluation framework. We have used our AR method in two use case scenarios to reconstruct the architecture of ARM-based Trusted execution environment and a Raspberry-pi platform have extensive application in embedded systems and IoT devices.

## 1 Introduction

The increasing complexity in hardware and software domains has resulted in an extensive complexity in computing systems. Use of a large number of SW/HW elements developed independently, multi-layered software stacks, huge system software with millions of lines of codes, motherboards with multiple chips having standalone processors, heterogeneous system architecture, emergence of new technological concepts (such as zones, fault domains, virtual sessions, isolated environments and etc.), the existing trend toward platform-based design [7], commodity chips and closed-source software blobs make most current computing devices even more sophisticated than before.

Achieving a high-level view of the system's component and their relationships -called the system's architecture - is a crucial requirement for the maintenance phase of computing systems. It is also beneficial for effective use of existing resources for a specific application. More important, providing the desired security and trust to the platform is also not possible without deep characterization of the system.

The need for discovering the architecture of a software application is an accepted topic in the software engineering domain and there are a plethora of techniques to reconstruct the architecture of a software application in the literature. These efforts have been also called architecture extraction [37], architecture recovery [25] [40], and reverse architecting [6]. Similarly, there are efforts in the field of hardware reverse engineering domain [15] [35].

Unfortunately, there is no well-known procedure for architecture extraction for a computing platform as a whole in the literature. We believe this is a missing link in the context of computing system analysis without which other reverse-engineering efforts would fail to cover some aspects of the system at SW/HW boundaries. In other words, combining software architecture and hardware architecture views separately would miss some SW/HW relations which are critical for understanding the system.

This HW/SW combined architecture reconstruction (AR) becomes more important when considering the close in- teraction between software and hardware components in today's computing ecosystem. Motivated by the technology, these interactions have multiple origins: First is the trend toward softening the hardware components [43] which can be seen in programmable hardware chips, reconfiguration facilities, architecture-aware compilers, and similar cases. Second is the trend toward implementing special computa- tions with high requirements of trust or computation power by the hardware. Hardware-assisted synchronization, vir- tualization, and security extensions are examples of the mentioned trend. As the result of this HW-SW coupling, one needs to consider the combinational architecture of hardware and software to realize the in-hand system.

It is worthy to note that in most cases component developers/manufactureres haven't a such a holistic view about the architecture of different platforms in which their product may be used. Hence it wouldn't be reasonable to trust the component developers about safety, security, and maintenance of the system. Instead, it should be possible for the user to analyze the in-hand system without full documentation and sources available.

In this paper, we have proposed a method for architec-ture reconstruction of computing platforms. We have used an ontology-based approach in order to handle diversity in different existing technologies. Ontology in the context of information science is defined as *"a highly structured system of concepts covering the processes, objects, and attributes of a domain in all of their pertinent complex relations, to the grain size determined by such considerations as the need of an application or computational complexity"*[1]. We believe that ontology is an appropriate tool for knowledge management of computing platform domain.

To this end, we have proposed a computing PLATform Ontology called *PLATOnt*. The ontology covers different hardware/software technologies exist in platforms with one of the four well-known CPU architectures (Intel/AMD x86, Intel/AMD x64, POWERPC, and ARM). Inspired by the Symphony software AR framework [44], we then have adapted the method to be applicable to computing platforms.Hence our contributions include:

- Developing an ontology describing HW/SW components of a computing system -called *PLATOnt*-, which is evaluated by a domain-independent framework called OntoQA [42] to show its effectiveness compared to other existing ontologies regarding multiple metrics regarding its classes, relations and instances.
- Generalizing an existing software architecture reconstruction method to be applicable to a whole computing system (having both HW/SW components). The method is then used in the conext of two use-cases to show its effectiveness for extracting the architecture of the ARM-based Trusted execution environment, as well as a Raspberry-pi platform.

The remainder of the paper is as follows: In section 2 we have reviewed previous research literature related to our work. Section 3 include explaination of our main idea including PLATOnt development process and our AR method. In section 4 we have shown the effectiveness of our method in the context of two case-studies. Finally, section 5 and 6 have been dedicated to discussion and conclusion respectively.

## 2 Related works

In this section we have briefly reviewed the existing efforts in the field of architecture reconstruction. Since we have proposed an ontology-based AR scheme, we have also enumerated the ontologies developed to describe computing platforms.

### 2.1 Software architecture reconstruction

Architecture reconstruction is a kind of reverse engineering effort in software engineering domain. While most of software AR efforts require the source-code, they may use other inputs such as run-time information [34], static information of application binary, or design artifacts.

The practice may be done in a bottom-up (architecture recovery) or top-down manner (architecture discovery) which either are supported by specific tools. In bottom-up tools ( .e.g. ARMIN [31], Rigi [21]) the analysis starts by analyzing the source code. The top-down approach however, begins by some abstract knowledge (about the requirements or architectural styles [11]) which is used to build a hypothesis. The source code is then used to verify the hypothesis [29]. Finally, hybrid approaches integrate both analysis techniques (.e.g. Symphony [22]). Nokia's Symphony [22] is a view-driven software AR reference framework which is generalized by our method to be applicable to a system as a whole.

There are surveys comparing some considerable software AR ideas. Nayyar and Shafique [29] tool-based survey and Ducasse and Pollet process-oriented taxonomy [11] are two examples of such surveys.

### 2.2 Computing platform ontologies

While there has been no ontology-based AR effort in the literature, developing ontologies to describe computing platforms is an approach can be found in several domains.

Considering computing platforms as a part of context in which an intelligent agent operates, there are multiple context ontologies [32] [19] [16] include a *platform* class to help agize their hardware and software environ- ment. This ontology-based realization is then used to reason about, adapt to, and collaborate in their operational context. One of the most considerable efforts among such ontologies has been developed by Preuveneers et al. [32]. This ontology includes general information about hardware and software components of devices may be involved in an ambient intelligent environment.

Cloud computing is the second domain in which several platform ontologies have been developed for enabling better resource annotation [27], service discovery [47], configuration assessment [9], and domain comprehension [46]. While most of sub-classes used to describe platforms in these ontologies are identical with non-cloud environments, there are also extra concentration on cloud service models (e.g. virtual appliance class [9]).

Model-driven software architecture is another domain in which some platform ontologies have been proposed [45]. In this context, platform ontology helps to convert a platform-Independent Model (PIM) of an application to a platform specific one.

There are also multiple other domains which platform ontologies have been shown to be beneficial. As an example, Dibowski et al. [10] proposed a four layer ontology for hardware device description. The layers start with domain specific terminological approach and end with specific manufacturer vocabulary. This device description ontology is then used for formal design and commissioning of modern building automation systems. Another example is the mobile operating system ontology [17] has been developed by Hasni et al. by federating two single ontologies of Symbian and android operating systems.

There are three main shortcomings in the above mentioned ontologies. First is that they are so brief and are not able to cover many components of the system (.e.g. they do not cover buses and chipsets). Second, they have too coarse-grained view for those components they can cover. Third is that they are biased toward some well-known implementations. As an example hypervisor is a platforms component always launched as a subclass of software class (like in [33]).

However, a hypervisor can be implemented by the firmware of a separated hardware chip (.e.g. POWER architecture), or inside the processor itself (.e.g. ARM monitor mode).

Another related research topic includes abstract ontological discussions [26] [12] about hardware and software as two core concepts of platform ontologies. While Moors philosophical view [26] does not accept ontological distinctions between the two, Duncan [12] discusses how almost all well-known definitions of hardware and software have some kinds of shortcoming to reflect an ontological distinction and presents a more exact definition of these two core concepts.

## 3 Ontology-based architecture reconstruction

In this section, after proposing a brief motivation in section 3.1, we will elaborate our developed ontology (section 3.2), and the proposed ontology-based AR method (section 3.3).

### 3.1 Motivation

When discussing the architecture of a specific software component of a computing platform, there are many software architecture analysis methods in the literature which enable the developer party to reconstruct and analyze the architecture of its product. These architecture analyses would help the developer to better provide maintainability, safety, and security in different phases of its life-cycle.

When considering the big picture of the platform, there is a system of hardware and software systems which have been separately analyzed. However, the relationships of these components together can result in a new collective identitiy and cause some kinds of challenges which may haven't been considered by a single developer party.

In addition, sometimes there is a legacy system with no/less such up-to-date maintenance services. Hence it must be possible to realize how the system works as the first requirement for maintaining the system.

System architecture reconstruction enables the end-user to find an abstract architectural view to a computing platform. Below we will enumerate some example questions which can be answered by such view.

- How true is the claim of having an open-source boot firmware for this specific platform? Is it possible to develop a free code for a probable existing firmware blob?
- How can the execution of a specific application get optimized in order to achieve better performance? Is there any more appropriate co-processor (.e.g. parallel co-processor) embedded in the platform for other purposes? Is there any choice other than the main memory to be used during the execution? How can these changes be applied to the execution environment of the application?
- How are the security and performance of the OS installed on this platform? To answer this question one should note that there is not a single well-known OS installed on the system. There are multiple standalone chips (.e.g. Security Co-processor, base-band modem, remote management engine, etc.) with their own OS inside the system.
- Is there any OS-independent hardware access vector to the applications?

Contrary to the popular belief, the above questions (and similar ones) cannot be easily answered. The primary step to answer the mentioned questions is to identify each of platform component correctly, whether that is implemented by hardware, firmware, or software. They should be then analyzed by a big picture containing the components and their relationship. This would be possible via architecture reconstruction method.

### 3.2 PLATOnt ontology

ISO/IEC/IEEE 42010 [4] standard has defined the architecture as the *fundamental concepts or properties of a system in its environment embodied in its elements, relationships, and in the principles of its design and evolution*. Hence in order to reconstruct the architecture, we require a knowledge management tool to help the user to discover these elements and relationships in a given platform.

Considering the definition of ontology, we think it best fits for this aim. An ontology *consists of a set of concepts and relations between these concepts, describing their various features and attributes* [8]. In continue we will elaborate on the development procedure of our ontology of computing platforms called *PLATOnt*.

#### 3.2.1 *requirements*

The first requirement is that PLATOnt should have the ability to generalize various existing technologies with diverse set of names to achieve their main functional concepts. This generalization may occur between technologies which primarily look fundamentally different. As an example existing ontologies defined operating system and hypervisor as two separated classes (concepts). This scheme is the result of the poor realization of system software concept which covers not only OS and hypervisor but also many types of codes run independent of the two (e.g. ARM SVC monitor code). Another type of the generalization occurs when similar functions have different HW/SW implementation. As an example in the existing ontologies a firmware Trusted Platform Module (fTPM) [36] cannot be modeled as they consider only hardware resources. This is while the ability to involve fTPM-based secure boot is required during platform security analysis.

The second requirement is that the ontology shouldn't have a bias on well-known architectures (.e.g. x86) which exist in almost all of the existing ontologies.

Finally, the third requirement is that PLATOnt should divide a commercial platform to the components in a manner which can be repeated by the end-user. Hence it should not rely on HW/SW source codes or similar artifacts which are not always available. At the same time, it should be able to use such knowledge resources if available.

#### 3.2.2 *Methodology*

In this section we will describe our methodology to achieve PLATOnt. Multiple ontology building methodologies have been proposed for different domains. We have followed the ontology development process proposed by Noy and McGuinness [30] as a well-known domain-independent method. Figure 1 shows the required steps and their application in the case of PLATOnt development.

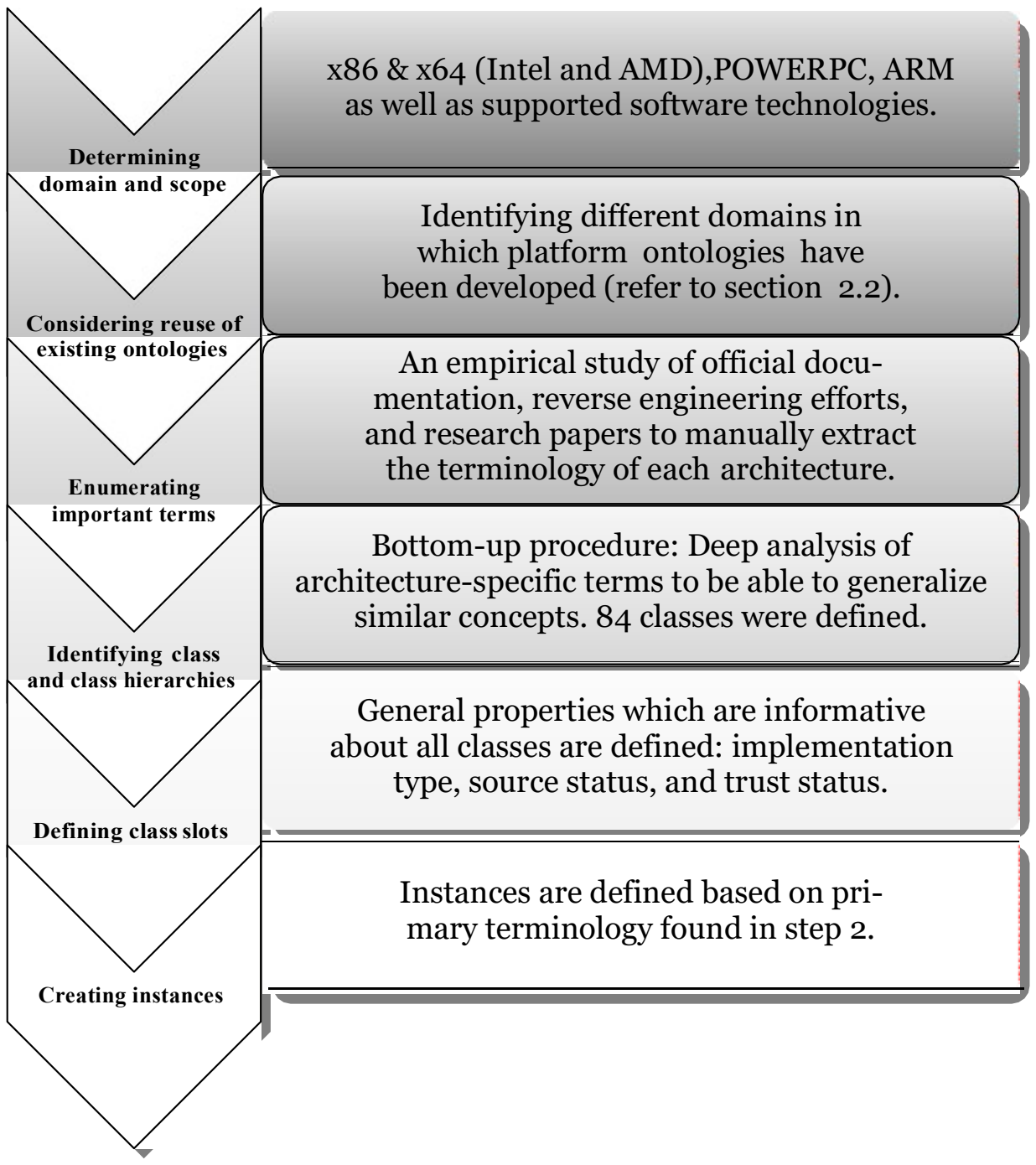

Fig. 1. Application of Noy-McGuinness ontology development method [30] for PLATOnt.

The first step is to determine the domain and scope of the ontology. We have limited our ontology to computing systems having five general purpose CPU architectures. These architectures are Intel/AMD X86 and X86-64, POWERPC, and ARM. Different software developed for these hardware architectures have been also considered inside the domain of the ontology.

We should then find and reuse probable existing ontologies. We have looked for such ontologies in different domains. As was mentioned in section 2, there are few platform ontologies in the literature with few classes (less than 20 classes) highly biased for X86 systems. So there were few reusable cases available.

The next step is to enumerate important terms. Most of the vocabularies used in the context of a platform are specific to a CPU architecture (which in turn affect other consistent HW/SW components). Extracting terminology of an architecture is not as easy as it looks. Our first effort was to use a text analysis application to automatically extract the related terminology for each hardware architecture. We have tried Tropes [3] terminology extraction option and applied it to the raw text of mentioned CPU manuals. However, we weren't successful since the terminology built by the software include a huge set of words which were mostly instructions, registers and their compositions[1]. As an example, only one of the volumes of Intel architecture manual produced more than 10000 terms which the majority was not usable to our ontology.

We believe that the reason for this failure is in the characteristics of current CPU documentations. Considering the Instruction Set Architecture (ISA), micro-architecture, and system design as three main disciplines of computer architecture [41], CPU documentations often address the first and second items. In fact, system design is not explained explicitly in any document and is expected to be realized within ISA explanations.

In addition, any CPU architecture can host different system software (including operating systems, and hypervisor), or can be placed on different motherboards with different devices and firmware. These all compelled us to manually extract HW/SW keywords for our target platforms.

At the next effort, we have tried to manually extract the terminologies. In order to ensure the finding of important terms we have used tens of papers on the topic of low-level attacks to the target architecture in addition to the official manuals. The attack background described in such kinds of papers include architecture specific terminologies which augment our initial findings. Similar effort has been performed to extract key terms from the documentation of operating systems and hypervisor products of each architecture.

The achieved set of words were mostly architecture/manufacturer specific and one needs to deeply analyze their functional characteristics to achieve their core concept. This was done through the third and main step which is to define the classes and the class hierarchy. Classes have been defined in a bottom-up manner. Generalizing concepts shared between different keywords, the core concepts have been set as an ancestor and more specific concepts became their sub-classes.

After defining the classes, we should define the properties and relationships. Hierarchical relationships (Is_a) are the most common relationship which exists between classes and their sub-classes. However, the more well-defined other kinds of relationship are defined in the ontology, the more informative the ontology will be compared to a simple taxonomy. [2].

In practice, since different kinds of non-hierarchical relationships between the components form different architectures, it is not possible to pre-define all relationships between classes and the user is asked to use the relations to describe the target system.

Finally, after the mentioned steps we should place instances under their regarding classes. Instances are those manufacture-specific keywords we have extracted in the second step.

### 3.2.3 Ontology

The final ontology is developed in the OWL format by Protege ontology editor [28]. The ontology has seven main super classes and set of three types of properties. In continue, we will briefly describe these main entities.

1. As an example *cmpicmpi8* is an instance term extracted from POWERPC manual.

2. In theory, properties can have restrictions called facets. Though we have defined facets, we think they are not much applicable in the current version of PLATOnt.

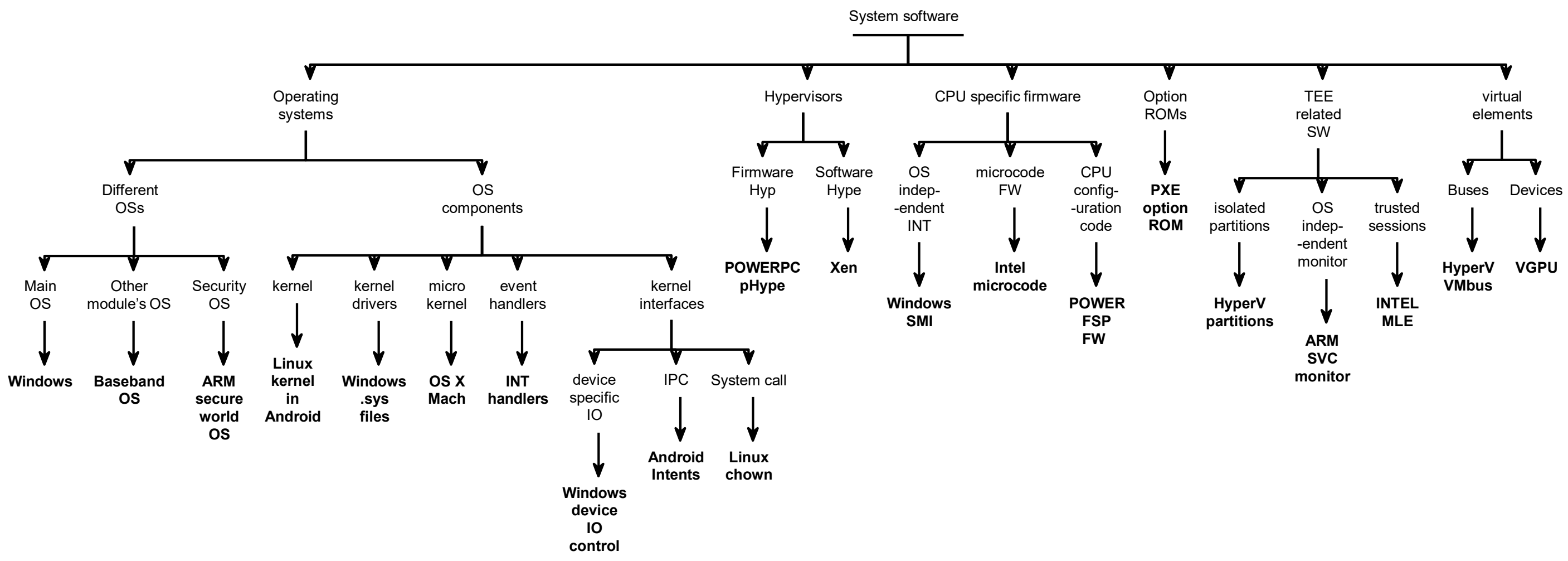

Fig. 2. System software class of the PLATOnt ontology. Bold Leaves are individuals for each class.

- **User-level software**: Considering different types of operating systems and application architecture we have defined seven types of user-level code including libraries ( .e.g. Windows DLL), user-level drivers, run-time environments, interpreters (.e.g. JVM) , virtual machines (.e.g. VMware workstation), application frameworks (.e.g. Carbon, Cocoa in OS X), and finally the applications.

- **System software**: This class includes codes which are run with higher privileges (ring/mode/level) of the CPU. There are various kinds of components we have considered under the system software class (figure 2):

- Operating systems are the first subclass. We have considered either different types of OSs as well as different sub-components of an OS in the PLATOnt. PLATOnt classify operating systems as the main CPUs OS or stand-alone chip's OS (e.g. base-band OS, TPM OS, Intel ME Minix OS or etc.). Today's modern operating systems have more complex software architecture which can be reflected in the ontology for further architectural analysis (e.g. Android or Mac OS X). Hence we have defined different components of an OS too. These components may be the kernel, microkernel (.e.g. Mach in OS X), Kernel drivers, and different types of kernel interfaces.
- Virtual machine monitors (.e.g. XEN), virtual buses and devices.
- Some types of system software components are implemented by firmware. Examples include firmware devices (.e.g. fTPM [36]) and option ROMs[3]. In addition CPU specific firmware components are another important kinds of system software.
- Trusted Execution Environments components: Today's platforms can support different kinds of TEEs. The software for these TEE environments have been enumerated under this class.

-**Boot components**: There are various types of system components involved in boot process. Other than the main boot firmware, there are next stage bootloaders with different functionalities such as hardware initialization, security, multi-boot, and diagnosis. Considering to utilities for providing a chain of trust, there is also a pre-authenticated code which works as the trusted primary component of the chain.

-**Processors**: processors are the next main class which covers the central processing unit (CPU) as well as other processors which are used in today's platforms. This includes device controllers and co-processors (.e.g. GPU , security co-processor). It is worthy to note that processors are not essentially hardware implemented. There may be logical processors supported by the CPU, or in virtual devices installed on the software stack.

-**Connectors, devices, and debug facilities**: There are many types of connectors in a platform. Chipsets, bridges, hubs, and physical buses are examples of hardware connectors. Different devices and debug facilities are defined as separated classes. Considering virtual environments, these facilities can be implemented by software as well.

Unlike all of the existing ontologies which use hardware and software as two main super-classes, we have found out these are implementation-related properties of components rather than the basic ontological concept. As stated by Tanenbaum [41], "hardware and software are logically equivalent" and "operation performed by software can also be built directly into the hardware and any instruction executed by the hardware can also be simulated by the software". A comprehensive discussion in [12] shows that it is hard to find an acceptable distinction criterion between hardware and software.

Softening the hardware components can be realized in the context of technologies like virtualization, reconfiguration, FPGA fabrics which are discussed in [43]. In the opposite direction, hardening some software components may be done to increase the computation power or trustability which is observable in hardware synchronizations (.e.g. hardware transactional memories), hardware-assisted security solutions (.e.g. TPM), and hardware virtualization CPU extensions.

As the result, PLATOnt includes different types of plat-

3. Option-ROMs belong to the system devices which are often loaded into main memory and executed by the CPU

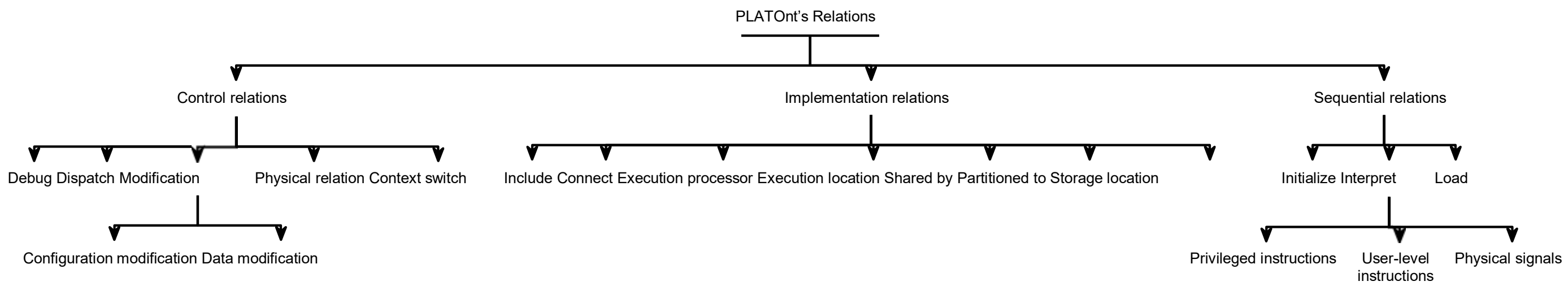

Fig. 3. General relations defined by PLATOnt between platform's components.

form components each have some general properties including implementation type (Hardware [4], software, or firmware), source status (closed, open, hybrid), and trust status (trusted, untrusted).

Due to space limitation, the full ontology cannot be placed within the paper. The OWL version of the ontology as well as its SVG visualizaion would be available on the PLATOnt webpage [5].

The classes can be found in almost all of platforms with different names. However, their properties and their relations differ between platforms. We have classified different relations exist between components into three following classes (Figure 3):

1) Control relations: Some components of the platform are designed to be able to have some controls over some others. This control operation may include *configuration*, *debugging*, *modification*, or different kinds of *execution redirection* (.e.g. interrupting, call-gating, and etc.). Sometimes the *physical* position of a component makes it possible to have some controls over other devices which it mediates the flow of information to them (physical man-in-the-middle).
2) Sequential relations: These relations indicate situations in which a component has some kinds of privilege because it can access some data belong to another component after/before the owner. These include *load* or *initialization* of the component, or *interpreting* some instructions the component cannot execute solely.
3) Implementation relations: There are some relations between component which regards to implementation of the platform. System resources may be *shared* by some consumer components, or may be *partitioned* to separated parts. Executable codes are *stored* in a storage medium, *loaded* by a loader component to a memory component and then are *executed* by a processor.

#### 3.2.4 Ontology Evaluation

In this section, we have evaluated the PLATOnt based on ONTOQA framework [42]. The framework is a domain-independent evaluation scheme and can be applied to any ontology. The scheme includes ten metrics regarding different aspects of the ontology (classes, relations, and instances).

4. When discussing hardware components we distinguish between ICs and SoCs (System on Chip).

Since ONTOQA can be used to compare two ontologies, we have also calculated the metrics for the platform ontology have been proposed by Preuveneers et al. [33] which was the largest ontology on computing platforms mentioned before in section 2. This was done automatically by giving the OWL codes of the ontologies to the ONTOQA open-source application.

Table 1 shows the metrics and their corresponding values. ONTOQA framework provides a weight for each metric, in order to indicate its importance. By considering these weights, a total score (weighted summation) can be assigned to each ontology which is more insightful for comparing different ontologies in the same domain. As the table suggests, number of defined classes, different types of non-hierarchical relationships, number of instances, and the average number of subclasses are the most important metrics that have caused a considerable increase in the PLATOnts score.

### 3.3 Architecture reconstruction

Inspired by Symphony software architecture reconstruction framework [44], our architecture reconstruction method includes three main steps by a same *extract – abstract – present* approach. Figure 4 shows these steps which are elaborated in the following subsections.

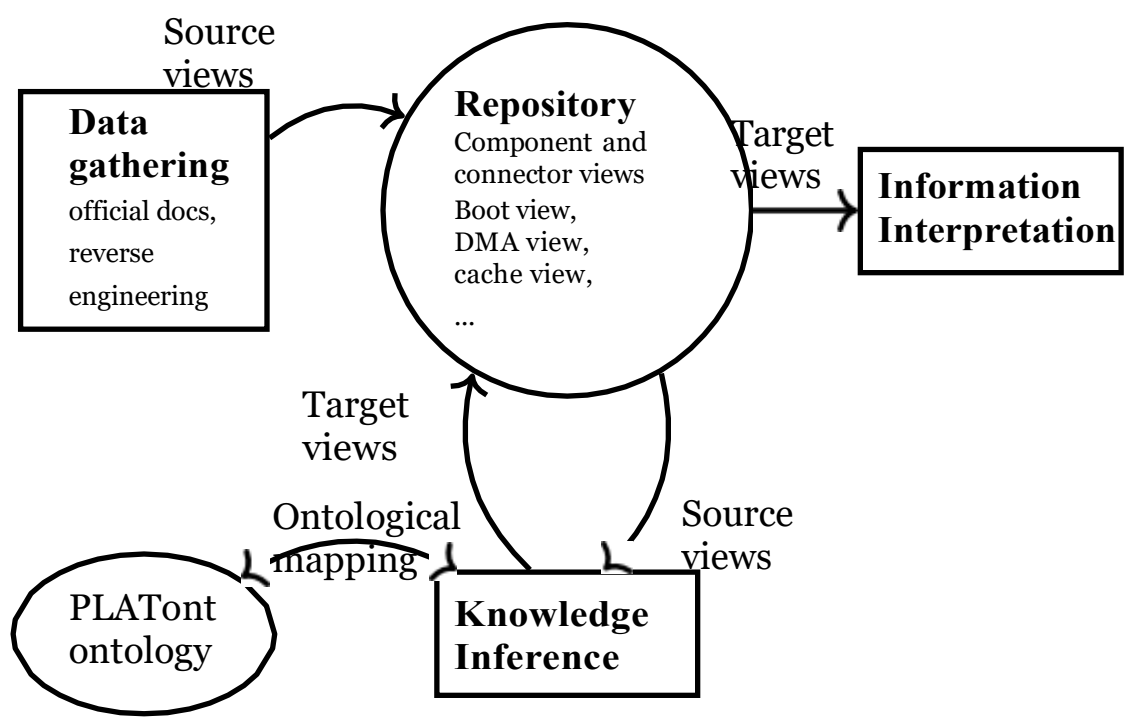

Fig. 4. Symphony-inspired ontology-based system architecture reconstruction.

#### 3.3.1 Data Gathering (extract)

In this step, the available artifacts for the platform should be deeply analyzed to recover architectural concepts (platform's elements, different kinds of communication mechanisms and relationships exist in the system).

TABLE 1
ONTOQA metrics [42] used to evaluate PLATont and their values(* values were not available in the reference paper.).

| Metric | Definition | Weight | PLATOnt | Preuveneers et al.[18] |
|---|---|---|---|---|
| Class count | Number of defined entities | 0.10 | 84 | 17 |
| Relationship count | Number of defined relationships | 0.07 | 33 | 3 |
| Relationship richness | Non inheritance relationships divided by the total number of relationships | 0.08 | 28.44 | 18.75 |
| Inheritance richness | The average number of subclasses per class | 0.08 | 4.15 | 3.25 |
| Average Class Height | The average height of classes | 0.06 | 2.45 | 3.17 |
| Attribute richness | Average number of attributes(slots) per class. | 0.03 | 0 | 0 |
| Tree balance | * | * | 1.94 | 1.3 |
| Individual count | Number of defined instances | 0.08 | 107 | 0 |
| Class Richness | The ratio of the number of non-empty classes (classes with instances) divided by the total number of classes defined in the ontology schema | 0.03 | 71.42 | 0 |
| Class Instance Coverage | * | * | 1.63 | 0 |
| Average population | The number of instances of relationship in the KB compared to the total number of property instances in the KB. | * | 1.27 | 0 |
| **Total score** | | | **24.16** | **3.86** |

AR can be performed at different levels of abstraction. The granularity of analysis should be such that it covers all available information to the end-user. The user is asked to find any execution unit inside the system regardless of its implementation type (software, firmware or hardware). For software and firmware components, the independently developed codes would be considered as separated execution units. For the hardware, any known subsystem which can be considered as a separated block with a specific functionality would be an execution unit (whether it is a separated IC or within a SoC).

This is important specially when having a system which its hardware has been designed by platform-based methodologies. This is because such *meeting-in-the-middle* designs - which are common approaches in embedded system design- result in more complex internal blocks each have their own processor, memory and etc.

At the first level of inspection, the components on main board of the system are identified. A general block diagram of hardware components is often exist in the manual of the board. Further search is required about IC numbers found in the previous step to continue the component inspection in a recursive manner. In this phase, different processors, controllers which may exist inside ICs, their capabilities, operational modes, interrupt types, hardware extensions, and similar information should be documented or tagged as unknown partitions. In the end, we expect to have nested diagrams of components which recursively build the main block-diagram of the system. After identifying hardware components, similar effort should be conformed to inspect software components of the system.

Since the goal of this phase is to gather as much data as possible about the system, lexical analysis can also be beneficial. Looking for some golden keywords help the analyzer to extract different components and relationships exist in the platform. Boot sequence, processor operating modes, transparent execution, and privilege levels are examples of such golden keywords. In any inspection step, if the intended information cannot be achieved it should be tagged as unknown and the inspection will continue in its next phases.

The data gathered in this phase form a source view of the platform. It should be noted that the architectural analysis has different views in software and hardware contexts. In the hardware context, architecture often involves a structural view, this is while behavioral views are more prevalent in the software context. When considering different inter-component relations at hardware-software boundary, in addition to the structural view, there are some relations at HW/SW boundary which forms some systems behavior. For example, consider the relation of an operating system with CPU and memory. The OS is loaded into the memory and is executed by the CPU. When considering multiple software execution units, there are also such behavioral relations between components.

### 3.3.2 *Knowledge Inference (abstract)*

As it is shown in figure 4, in this step the achieved source view about the platform should be converted to a desired target view.

Each target view reflects the system architecture from the perspective of a related set of concerns [20]. While standard viewpoints [24] exist for software AR efforts, our method has a holistic viewpoint which considers either hardware and software components sometimes referred to as *components and connector*, process, or *execution* view in the software architecture literature.

Choosing such a reference viewpoint, custom viewpoints can be derived based on the specific goal of the reconstruction. Boot view (.i.e. reconstructing HW/SW components involved in the boot procedure) and DMA view (.i.e. reconstructing components involved in direct memory access operations) are examples of such views. Inspired by Symphony, these views are extracted and stored in a repository to be available for different analysis purposes.

In the Symphony software AR framework "*the mapping rules and domain knowledge are used to define a map between the source and target view*" [44]. Unlike Symphony, the mapping from the available source view (any kind of documentation

```
procedure ONTOLOGY-BASED MAPPING PROCEDURE
    Input: Source view, PLATOnt ontology, the desired target view
    Output: Component and connection view (probably partial)
                                                d Find probably overlooked components
    for all classes (C ) in PLATOnt do
        if no entity E exists with same a concept with C then
            Review documents to find a probable instantiation of C.
        end if
    end for        d Find mapping for key words found in Data gathering phase
    for all entities (E) in source view do
        if E exists in PLATOnt instances (I) then
            Component(E) = PLATOntclass(I)
        else if C exists in PLATOnt classes with same concept as E then:
then
            Component(E) = C
        else
            Update the PLATOnt.
        end if
    end for
                                                d Find connections
    for all found Components do
        Define relevant connections with other components from PLATOnt
relations.
    end for
                            d Check which components are involved in the target view
    for all found Components do
        if Component is involved in the functionality of the desired view
then
            Add Component's connections which are related to the
Output components to Output connections .
            Add Component to the Output components.
        end if
end for
end procedure
```

algorithm 1: Pseudo code for ontology-based mapping procedure.

achieved from the previous step) is not rule-based in our method. Here we use an ontology-based mapping scheme.

Algorithm 1 shows the psuedocode for this ontology-based mapping which is performed manually. The analyzer is asked to use the PLATOnt ontology to make sure that all components of the platform have been identified. After this pre-processing phase, PLATOnt relations are used to connect the identified components.

Due to the complexity of the extract-abstract process, the user may require to return back to data gathering artifacts to find individuals for some classes/relations which were not considered before.

#### *3.3.3 Information Interpretation (present)*

Finally, the target view provides by the method should be used to provide an analytical report for the reconstruction stakeholder, such that it can be used to solve the initial problem.

As it will be shown in section 5, the initial problem can be proposed in different domain, including free software development, maintainability, execution optimization, forensics, and security.

The architecture should be also visualized in this phase. This visualization is an important evidence attached to the final report which is the main outcome of this phase.

One choice is to use a layered view of the system. Though useful, layered diagrams cannot show various types of relationship between component. In addition, almost all of the existing platforms do not have a fully layered architecture. The use of such representation ignore those relations violate the layered scheme. More important, layered scheme does not match the components and connection view of our method.

Systems Modeling Language (SysML) [14] is another choice which is an extension to UML software modeling language. It has added 9 new types of diagrams to UML in order to reduce its software-centric restrictions such that it can be used for modeling systems.

We will show how can this architecture reconstruction can be applied through two case studies.

## 4 Architecture Reconstruction Methodology Evaluation

In this section, we have shown the effectiveness of our proposed method through two case studies. The first case targets the reconstruction of ARM Trusted Execution Environment (TEE) [13]. This recovered architecture is then used in the context of AR of a Raspberry-pi single board computer platform.

### 4.1 Partial architecture reconstruction

Architecture reconstruction can be used for reconstructing a special part of a computing platform. In this case, the analyzer is asked to follow the method to obtain partial architecture of the system including all components and their relationships which are involved for the target functionality. Partial realization can be beneficial to compare different technologies with a same functionality as well as similar systems with partial difference.

In this section we have focused on the reconstructing the architecture of ARM TEE [13] available in millions of devices. In addition, it is used by AMD secure technology and Intel Management engine SoCs.

ARM-based TEE AR is performed based on the ARM official documentations for ARM Cortex-A (64bit). Cortex-A family of ARM cores includes application processors equipped with TrustZone security extension. In practice, this basic specification will have some changes in ARM-based chips manufactured by different OEMs.

In a Tustzone-enabled ARM platform, hardware resources are designed to support two separated execution environments -called secure and normal worlds- for running untrusted and trusted executables independently. The critical code executes on the secure world which the normal application can communicate with.

Data gathering phase is done by collecting official documentation as well as different ARM-based attacks and defenses proposed in the academic literature. The later include over 180 papers which could help us to better understand the architecture. Knowledge inference phase was done by the help of PLATOnt and we have found 32 components involved in ARM-based TEE. Table 2 shows how our reconstruction method models different facilities involved in an ARM-based TEE and mapping found elements to the PLATOnt classes.

The security context of the processor cores (secure and normal worlds) is modeled by two set of virtual cores. This is because according to the official documentation, "Each physical processor core provides two virtual cores. one considered non-secure and the other secure, plus a mechanism to robustly context switch between them" [2]. As the table suggests, we have used two separated components belong to the two set of virtual cores executing the software stack dedicated for each world (running in a time-sliced fashion).

TABLE 2
Description of how the method reconstructs TrustZone-Aware (TZA) elements.

| Type | Platform property | Mapping to PLATOnt classes |
|---|---|---|
| Resources | TZA RAM/ROM/Flash storage<br>TZA L2 cache | Two separated *memory device* components partitioning the main device.<br>Two separate set of *cache* components sharing lines of main cache component. |
| processor | Processor security contexts<br>CP15SDISABLE input signals<br>Interrupt requests (FIQ/IRQ)<br><br>Monitor mode<br>Hype mode | Two set of virtual *cores* .<br>lock-down node as a *system software* component with modification access to the CPU.<br>Two separated *system software* components for the interrupt handler codes with their own *interrupt controller* which normally deliver the interrupts to the monitor code.<br>*system software* component with context switch access to the CPU.<br>system software component which *interprets* privileged instructions of the normal world's OS. |

Memory components (ROM/RAM) are partitioned to two separated parts for each security contexts. Hence we have modeled them by adding these pairs as new compo- nents. The L2 cache is handled in a different manner which does not follow partitioned layout. In fact, the cache lines are shared by the two security contexts and are tagged by a NS bit indicating the context which the data belongs to. It is worthy to note that unlike the memory, secure world context is not allowed to access lines belong to the normal world. We have dedicated two set of separated components for different memory types of worlds.

Trustzone TEE can be integrated also with traditional system virtualization (in its normal world). Hype mode is a separated CPU mode supports trap-and-emulate required for virtualization in the non-secure security context. This mode is modeled by a Hype code component which interprets privileged instructions of the normal OS.

Another ARM CPU mode is the secure monitor mode. There is a trusted monitor (executes on this mode) code which is executed in the context of secure world and is responsible for context switching the processor between two worlds. Calling the Secure Monitor Code (SMC), a privileged system software can request for the context switch to secure world. The monitor is also the original code can trap different interrupts. ARM architecture has defined fast interrupts (FIQ) as well as normal IRQs which are recommended to be deployed for separating interrupts of the two worlds. These interrupts are delivered to the ARM core by Trust Zone Interrupt controller (TZIC) SoC which includes two separate interrupt controllers shown in figure 5. Interrupts are then handled by the monitor (or the system software components which the monitor delegates the handler).

A less discussed configuration facility in ARM architecture is the lock down mechanism. According to the official documentation *"Systems that want an additional level of protection can use a signal input into the processor core to lock-down some of the critical Secure world configuration options in CP15"*. To this end, the CP15SDISABLE processor input signal should be configured at boot time *"before passing control to the Normal world software"* [2]. We have also mentioned this option in the reconstructed architecture by a component for indicating lock down configuration code.

In the third step, we have visualized the reconstructed architecture ( Figure 5) to be applicable for different interpretation contexts.

The achieved component and connection view can be analyzed through different points of views. Since the Trustzone extension inherently has been developed to provide hardware-assisted isolation, we have interpreted the recovered architecture from the isolation point of view.

As the figures suggest, most of the involved components can be disparted into two secure and normal contexts (shown by gray and white nodes). Despite, there are some lines between the two contexts. Hence, without accurate characterization of the communication between these isolated contexts, efforts for the use of ARM TEE may fail and provide a legitimate exploitation path for an attacker.

Considering the secure world as the more privileged context, one can classify the cross world edges into two sets: privileged to normal, and normal to privileged. Lock down code's ability to change the normal worlds access rights, context switch of the normal world's virtual cores, and modification access of secure world executables to the memory dedicated to the normal world are edges include in the former set. An accurate study of such edges is critical to evaluate the security of the provided TEE.

As an example, secure world software stack has the modification access to the memory partitions dedicated for normal worlds. This is while the cache lines do not follow the same rule and normal world lines are not accessible through the secure world. This is what was exploited by cachekit [48] rootkit which hides a malicious program resides in the normal world's cache from any monitor/forensic tool running in both worlds.

The other set of probable cross world edges which should be analyzed are those in reverse direction. Since the TEE designers assume the lower privilege of the normal world, it is expected that the system has no such edge. Although there have been some SMC call vulnerabilities [39] [38] which allows an attacker to access the secure world's memory. These modifications, however, are not architecturally defined in the system and made possible by the insecure implementation of monitor code.

### 4.2 Whole system architecture reconstruction

In this section, we have shown how our architecture reconstruction method is applicable to a specific platform as a whole. We have chosen Raspberry-Pi (model A) system as the target which is used by many embedded devices.

#### 4.2.1 Data Gathering and knowledge inference

In the first phase (data gathering) we have started with Raspberry-pi official documentations and extracted few HW components: Broadcom BCM2835 (CPU & GPU), 256/512MB SDRAM, USB/Ethernet/HDMI/Audio ports, SD Card Slot, and the GPIO. We then have gone forward by the practices mentioned in section 3.3.1. We have tried to find documentations belong to each of the components by finding their part number or exact chip name. Since the main part of the functionalities is embedded inside the proprietary BCM2835, we have tried to find out as much as possible about the SoC. In the case of SW components, we have found different existing software codes applicable to Raspberry-pi including boot utilities, hypervisor, and etc.
Next, we have used PLATOnt in the knowledge inference phase. To this end, we have tried to map each concept exist in the ontology to a found keyword from the Raspberry-pi. This makes it more straightforward procedure of documentation analysis.
A more challenging task was to find the relations exist between the components. The master-slave relation between Videocore GPU and the ARM processor has added more complex relations between components. There are multiple communication mechanisms between ARM and Videocore. Providing a shared memory space is the first one which can be used for zero-copy operations. The second mechanism is to use interrupts. Since these processors use a shared memory on a physical bus, the GPU (master) should be able to interrupt the operation on the ARM when needed.

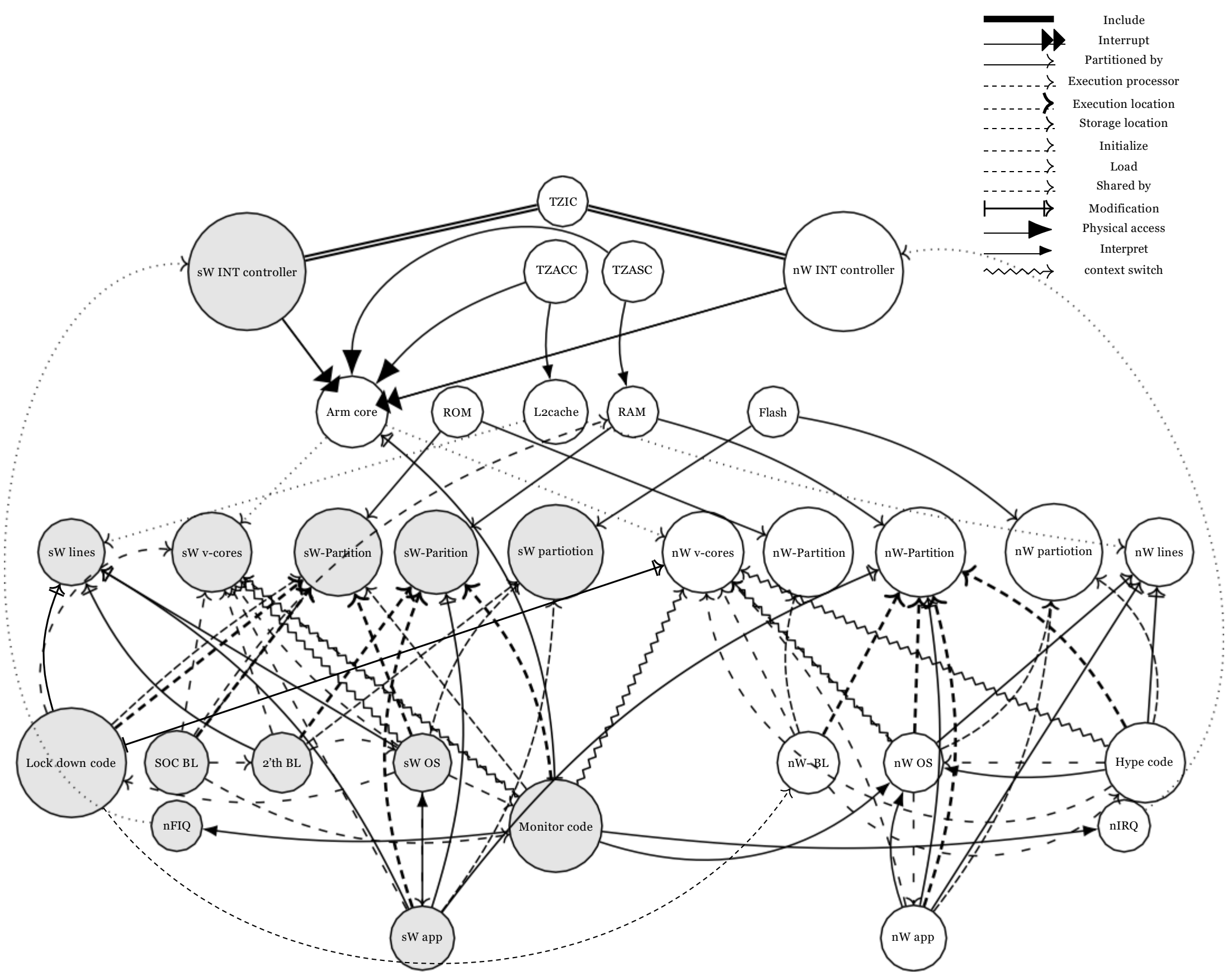

Fig. 5. ARM Trustzone-based TEE architecture

#### 4.2.2 *Information interpretation*

We have used SysML graphical language to represent the architecture by the desired component and connection view. We have made the complete diagram (which includes 50 components densely connected to each other) on our webpage [5].

The architecture indicates the Raspberry Pi platform is not as simple as it may look in the first glance. There are more than 6 processors, three different operating systems[5], more than four stages of bootloaders, and binary blobs inside this popular platform.

There are multiple viewpoints which can be derived from the reference achieved view. Table 3 shows some of these viewpoints and a brief output report considering a specific concern by the architectural view.

## 5 Discussion

Providing an ontology of computing platforms can be beneficial in multiple applications. In general, such an ontology can augment existing subject classification systems like ACM Computing Classification System (CCS) [1] or Mathematics Subject Classification (MSC).

5. We have assumed the memory card to use 8051 RAM controller. If an ARM-based memory is used there would be one/two other operating systems in the Raspberry platform.

TABLE 3
Information interpretation phase applied to Raspberry-pi AR

| Initial problem | Target view | Brief Interpretation |
|---|---|---|
| Free bootloader development | Boot view | Three primary boot components run on the GPU (proprietary processor instructions) and then three components execute on the ARM processor. These components have been stored in different storage components and have different execution locations (ROM, Cache-as-RAM, RAM). Developing code for each stage requires accurate study of tasks should done by that specific stage, the ISA for its execution processor, and a (HW/SW) mechanism to update the storage media. |
| Boot security | Boot view | Starting the boot sequence by the GPU instead of the ARM application processor, will make it more difficult for a malicious application (bootkits) to inject a stealth code into either secure and normal security contexts. In fact, the GPU can be a more secure location for positioning monitor code for checking the boot sequence. Reverse engineering efforts have made it possible to run arbitrary code on the GPU. Despite, such a monitor would be vulnerable to possible attack vectors against the less known closed-source GPU processor (Which has its own Minix-based OS and Alphamosaic processor). |
| Optimizing deep learning application execution on Raspberry (GPU programming) | GPU view | The GPU consists a Vector Processing Unit (VPU), the DMA engine, and 3D pipeline. A Quad Processing Unit (QPU) is in the heart of the 3D pipeline which can execute parallel computations. The VPU (a Dual core Alphamosaic processor) offloads OpenGL commands when executing applications running on the ARM. However, the VPU's 3D pipeline is fed by a hardware manner and commands are executed by the QPU. Hence the more knowledge about the QPU position in the system architecture, the more parallelization would be possible by providing accurate compilers. |
| Memory forensic tool development | memory view | A single SDRAM is partitioned multiple times ( .e.g. GPU/CPU, ARM secure/normal world, and kernel/user partitions) and there are different components of the system with their specific access rights to some of these partitions. Memory addresses are translated via multi-stage procedure. Hence it is important to consider which partition is accessible via the location in which the tool for dumping the memory is located? By considering the shared SDRAM between GPU and the ARM, one should consider the effectiveness of the TEE protections supported by the Trustzone in this architecture. The original access control mechanism applied by the ARM on the two RAM sections dedicated to the two security contexts cannot prevent the GPU access to the secure world RAM. Hence the question is about whether or not another mechanism defined by the Broadcom in order to protect RAM section dedicated to the secure world? and how much such probable protection is effective. Without answering the question, the TEE facility of the ARM processor looks to be futile in the Raspberry architecture. |
| Cache side channel attacks | cache view | While the ARM CPU and the Videocore GPU share most of system resources (internal ROM, SDRAM, and the storage), The L2 cache is only used by the GPU in the default setting (for better performance). The ARM processor, however, can enable sharing of the cache. The outcome of this cache sharing should then be inspected more accurate against different cache attacks. |

It should be noted that our method aims to help the user to handle the complexity of different technologies in a given system and reflect platform's component based on declaration of the manufacturer of the components. Hence it is not expected to be able to indicate probable hardware/software implants. In fact our method is a procedure to analyze the platform's inherent architecture. This would be beneficial when comparing different architectures, best use of embedded resources, alternating platform's binary blobs, locating best platform points for security products, and similar applications.

The input information required for our method may be achieved by official documentations, reverse engineering efforts, open architecture usage. In its ideal form, augmented ISA contract [18] [23] can result in richer input and output result of the proposed method.

Although PLATOnt has been tried to cover as much platforms as possible, it is hard to guarantee its completeness at any given time. The most updated version will be available on the PLATOnt web page [5] for further analysis by the community.

Another point is that the electronic chips with no soft code (.e.g. CMOS backup battery, CPU fan) are not considered in PLATOnt deliberately. That's because they do not contain any programmable part.

There are more systematic choices for interpretation step for architecture construction. Use of architecture description languages (ADLs) can be a choice. However, most of the existing ADLs cansnot cover HW and SW components simultaneously. In addition, they are developed for a custom goal which cannot simply used for other purposes. AADL is a standard language with HW/SW coverage which can be also extended to include the PLATOnt concepts. Hence we are going to implement the PLATOnt by AADL in the future.

## 6 Conclusion

In this paper we have proposed an ontology-based method for platform architecture reconstruction. Our method is inspired by Symphony [44] software architecture reconstruction framework and is an effort toward adapting Symphony to be applicable for a whole system.

Due to the insufficient ontological approaches exist in the context of computing platforms, we have developed PLATOnt ontology by using the Noy-McGuinness ontology development methodology [30]. Using metrics proposed in ONTOQA [42] we have evaluated the classes, relations and instances of our ontology and compared the results with an existing ontology to show our contribution.

We then have shown how can our method be used to reconstruct the architecture of ARM Trutzone-based TEE (partial architecture) as well as a Raspberry-pi platform (whole system architecture) as two case-studies.

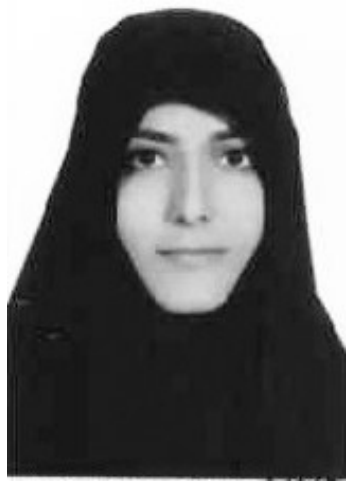

**S. Atefeh Musavi** received her B.S. degree in computer engineering from the Amirkabir University of Technology, Tehran, Iran in 2011 and her M.S. degree in information technology from the Sharif University of Technology, Tehran, Iran in 2013. She is currently pursuing her PhD in Tehran university. Her research interests include system security and information forensics.

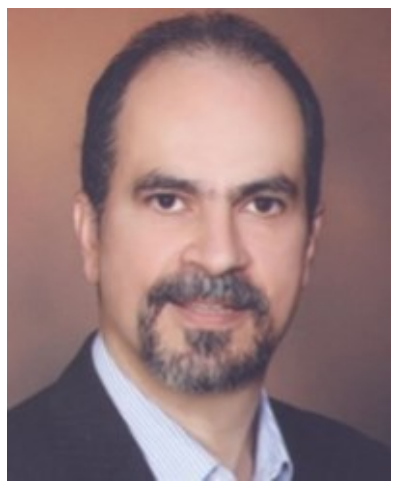

**Mahmoud Reza Hashemi** received his B.Sc., and M.Sc. in Electrical Engineering from the University of Tehran. He pursued his Ph.D. at the University of Ottawa, Canada. He is currently an Associate Professor at the School of Electrical and Computer Engineering of the University of Tehran. Dr. Hashemi is the co-founder and Director of the Multimedia Processing Laboratory (MPL). His research interests include reconfigurable hardware architectures for multimedia processing, computer architecture, multimedia systems and networking, and recently cloud gaming.